\documentstyle[11pt,psfig]{article} 
\begin{document}

\title{Gravitational lensing as folds in the sky}

\author{Silvia Mollerach and Esteban Roulet\\
Departamento de F\'\i sica\\
Universidad Nacional de La Plata\\
CC 67, 1900, La Plata, Argentina\\
email: mollerac@venus.fisica.unlp.edu.ar,
roulet@venus.fisica.unlp.edu.ar}

\maketitle

\begin{abstract}
We revisit the gravitational lensing phenomenon using a new
visualization technique. It consists in projecting the observers sky
into the source plane, what gives rise to a folded and stretched 
surface. This provides a clear graphical tool to visualize some
interesting well-known effects, such as the development of multiple
images of a source, the structure of the caustic curves, the parity of 
the images and their  magnification as a function of the source position. 

\end{abstract}

\section{Introduction}

The gravitational bending of light gives rise to many exciting
observable effects such as the multiple images of distant quasars, 
the  distortion of faint background galaxies by foreground clusters or
the microlensing of stars in the Milky Way and nearby galaxies. The
different phenomena that
gravitational lensing can produce have been widely studied and the
field has developed to become a major tool to investigate several
astrophysical problems, such as the determination of the Hubble
constant, of the amount of dark matter in the Universe, of the amount
of compact objects in the Milky Way, the search of planetary systems,
etc. \cite{bl92,sc92,na96,pa96,ro97,wa98}.

When the lensing mass distribution is concentrated at some distance
from the observer, e.g. if it belongs to a cluster along the line of
sight to a quasar or if it is a compact astrophysical object (possibly
a binary) along the line of sight to a star, one can talk about the
thin lens approximation, in which the `lens plane' is well defined.
A vector lens equation describes the position of the images of a point
source that is viewed through a distribution of matter acting as
lens. When this mapping is not single valued, multiple images of the
source arise, this is the so called strong lensing regime. 
 For non-singular lenses, images appear and disappear in pairs  
\cite{bu81}
as the source crosses lines in the source plane called caustics. 
These caustics hence  divide regions in the sky 
with different number of images. The images that appear or disappear
as the source crosses a caustic have opposite parities.

Here we describe a method that allows to visualize these properties in
a simple way and graphically determine the position of the caustics
as well as the number of images for a given lens configuration.
This consists in projecting the observer's sky into the source plane using
the lens equation. The surface describing the sky seen at Earth (the
`sky sheet') appears folded (if strong lensing occurs) and stretched  in
the source plane. The location of the foldings corresponds to the
position of the caustics, along which the magnification of the images 
 diverges. 
Every fold produces a pair of new images of the source with opposite
parities. Foldings can be closed lines or they can end in a point
called a cusp, where two foldings merge so that the sky sheet
unfolds. If the lenses are singular,
e. g. in the limit of point-like masses relevant for microlensing,
a tear appears in the sky sheet and one of the new images is lost.
This will be described in section 2 which is devoted to lensing by a
single lens. For single lenses the circular symmetry produces a
point like caustic when the lens is perfectly aligned with the
source. We also describe in section 2 how this caustic is transformed
when an external shear is added.
Some binary lens cases are then analyzed in section 3.


\section{The single lens}

\subsection{The lens equation}

As a first example we consider a single lens, corresponding to a mass
distribution located at a distance $D_{ol}$ from the observer, acting as
lens for a source located at a larger distance $D_{os}$ (for cosmological 
distances the angular diameter distance has to be used).  According to
general  relativity, light rays traversing the lens plane are deflected
by an angle $\vec{\hat \alpha}$ that depends in general on the distance to
the mass distribution. Thus, the images of a source located at an angular
position $\vec \beta$ will appear at angular position(s) $\vec \theta$
satisfying the lens equation 
\begin{equation}
\vec\beta=\vec\theta - \vec\alpha (\vec\theta),\label{leq}
\end{equation}
where $\vec\alpha \equiv \vec{\hat\alpha} D_{ls}/D_{os}$ 
and $D_{ls}$ is the distance from the deflector to the source.

For a point mass lens, which is a good approximation for star
lenses, the deflection angle is given by  
\begin{equation}
\hat\alpha = \frac{4 G M}{c^2 \xi}, \label{pm}
\end{equation}
with $\xi$ being 
the distance to the lens from the point at which the light ray
intersects the lens plane. The deflection angle points in the
direction from that point to the lens. In terms of the Einstein angle,
$\theta_E \equiv (4 G M D_{ls}/c^2 D_{ol} D_{os})^{1/2}$, 
the lens equation reduces in this case to 
\begin{equation}
\vec\beta = \vec\theta \left(1-\frac{\theta_E^2}{\theta^2}\right),
\label{lepm} 
\end{equation}
where  the angles are measured with respect to the direction towards
 the lens.
This equation has two solutions for every value of $\vec\beta$, except
for $\vec\beta =0$ (perfect source--lens alignment) 
where all the points along the circle
$|\vec\theta| =\theta_E$ (Einstein's ring) are images.

When the object acting as lens is a galaxy or a cluster, an extended
mass distribution has to be considered\footnote{Non-compact star-like
lenses have also been considered as possible responsible for
microlensing events \cite{gu96,za96}.}. Different models have been
proposed for this extended distribution
\cite{ha99}. A simple and widely used one,
that reproduces the observed flat rotation curves of galaxies, 
is the singular
isothermal sphere, which has a mass distribution given by $\rho(r)=
\sigma^2/2\pi G r^2$. Here $\sigma$ is the one-dimensional velocity
dispersion, which is related to the rotational 
velocity by $v_r=\sqrt{2}\sigma$. 
This model leads to a deflection angle
\begin{equation}
\hat\alpha = 4\pi \frac{\sigma^2}{c^2},
\label{sis}
\end{equation}
which is independent of the distance to the lens and points towards
the center of the mass distribution. The Einstein angle in this case
is given by
\begin{equation}
\theta_E = 4\pi \frac{\sigma^2}{c^2}\frac{D_{ls}}{D_{os}},
\label{easis}
\end{equation}
and hence the lens equation can now be written as
\begin{equation}
\vec\beta = \vec\theta \left(1-\frac{\theta_E}{\theta}\right).
\label{lesis} 
\end{equation}
A source located inside the Einstein ring will have two
images, one at an angle larger than $\theta_E$ and the
other one at a smaller angle. If the source lies outside the
Einstein ring it has just one image, while if it is aligned with the
center of the distribution the image will be  the Einstein ring.

The singularity in the center of the mass distribution can be smoothed
by a core region with finite density
given by 
\begin{equation}
\rho(r)= \frac{\sigma^2}{2\pi G r_c^2} \frac{3+(r/r_c)^2}
{[1+(r/r_c)^2]^2},
\end{equation} 
the deflection angle becomes
\begin{equation}
\vec{\hat\alpha} = 4\pi \frac{\sigma^2}{c^2}\frac{\vec\theta}
{\sqrt{\theta^2+\theta_c^2}} ,
\label{is}
\end{equation}
where $\theta_c\equiv r_c/D_{ol}$.
The lens equation can be written in this case as
\begin{equation}
\vec\beta = \vec\theta \left(1-\sqrt{\frac{\theta_E^2+\theta_c^2}
{\theta^2+\theta_c^2}}\right),
\label{leis} 
\end{equation}
where the Einstein angle is 
\begin{equation}
\theta_E =\sqrt{ 16\pi^2 \frac{\sigma^4}{c^4}\frac{D_{ls}^2}{D_{os}^2}-
\theta_c^2}.\label{einsis}
\end{equation}

This softened isothermal sphere has smoother behavior than the
singular models and hence we will often adopt it as a reference
model. In particular, it will be useful to discuss the formation of
image pairs. The analytical solutions of the lens equation 
giving the position of the image(s) for a given source is
trivial in the singular cases (eqns. (\ref{lepm}) and (\ref{lesis})), 
but to obtain $\theta_i(\beta)$ (where $i$ labels the images) 
in the non--singular case requires to find the solution  of the 
fourth order equation
\begin{equation} 
\theta^4-2\beta \theta^3+(\beta^2-\theta_E^2) \theta^2-
2\theta_c^2\beta\theta+\beta^2\theta_c^2 =0,
\label{poli}
\end{equation}
with the constraint sign$(\theta-\beta)=\ $sign$(\theta)$. 
This last constraint eliminates one spurious solution of
eq. (\ref{poli}),
leaving only one or three physical solutions which correspond to the
location of the images of the source.

\subsection{The sky sheet}

We turn now to the construction of the sky sheet corresponding to a 
given lens configuration. This is done  by considering  light rays 
which arrive to the observer along directions forming a regular
grid (in $\vec\theta$) and  following them back to the source plane 
according to the 
lens equation (\ref{leq}). The image of this grid will define a new 
surface in the source plane, which will be 
 stretched and folded. This problem is two dimensional, i.e. it involves
only the observers coordinates $(\theta_x,\theta_y)$ or the source
coordinates $(\beta_x, \beta_y$). However, to visualize the folds in
the sky sheet we have assigned also a vertical
coordinate $z$ to the mapping from the observer to the source skies. We
have conveniently chosen this vertical coordinate as being 
proportional to
the time delay between the actual path of the light of each image of
a source located at $(\beta_x, \beta_y$) and the straight
path from a source located at $\vec\beta=0$ in the absence of lenses.

The time delay for light rays coming from a given source due to the
presence of a lens has a geometrical contribution coming from the
extra path length of the deflected ray, plus a gravitational
contribution due to the time stretching in the gravitational
potential
\begin{equation}
t(\theta)=\frac{(1+z_l)}{c}\frac{D_{ol} D_{os}}{D_{ls}}\left(\frac{1}{2}
\alpha^2(\vec\theta)-\Psi(\vec\theta)\right),
\label{td}
\end{equation}
where we allowed for a finite redshift $z_l$ of the lens and 
$\Psi(\vec\theta)$ is proportional to the projected 
gravitational potential of the lens on the lens plane,
\begin{equation}
\Psi(\vec\theta)=\frac{D_{ls}}{D_{ol} D_{os}} \frac{2}{c^2}\int
\Phi(D_{ol}\vec\theta,z) {\rm d}z,
\end{equation}
with $\Phi$ the three-dimensional gravitational potential. $\Psi$
is scaled in such a way that the deflection angle is given by $\vec\alpha
(\vec\theta)=\vec\nabla_\theta \Psi$.\footnote{For the point-like 
mass $\Psi(\vec\theta)= \theta_E^2 \ln (|\vec\theta|/\theta_E)$, for the 
singular isothermal sphere, $\Psi(\vec\theta)= |\vec\theta|\theta_E$,
and for the softened isothermal sphere, $\Psi(\vec\theta)= \sqrt{
(\theta^2+\theta_c^2)(\theta_E^2+\theta_c^2)}$}
This time delay is measured with respect to the 
arrival time from the same source in the absence of lenses. 
To better visualize the sky sheets we have found more convenient 
to use the time delay with respect to a source located at 
$\vec\beta=0$ as the $z$ coordinate\footnote{Since the 
additional delay is
independent of the lens distance while the usual time delay depends on
it, for the plots we took for definiteness $D_{ol} = D_{ls}$.}.
Taking the $z$ coordinate proportional to the time delay allows 
to display the arrival time
ordering of the flux variations of the source for the different
images, so that flux changes will first appear
for the image with smaller $z$.

Notice that when the time delay is plotted as a function 
 of $(\theta_x,\theta_y)$ for a fixed source position $\beta$, 
a surface of time delays is obtained whose extrema 
(maxima, minima and saddle points) give the position of the images 
of the source according to Fermat's principle \cite{bl86}. 
This surface should not be mistaken with our sky sheet, which is
plotted as a function of $(\beta_x,\beta_y)$, i.e. for different
source locations, and the height $z$ is associated with the time 
delay of the actual images.

   \begin{figure*}
\vspace{0cm}
\hspace{0cm}\psfig{figure=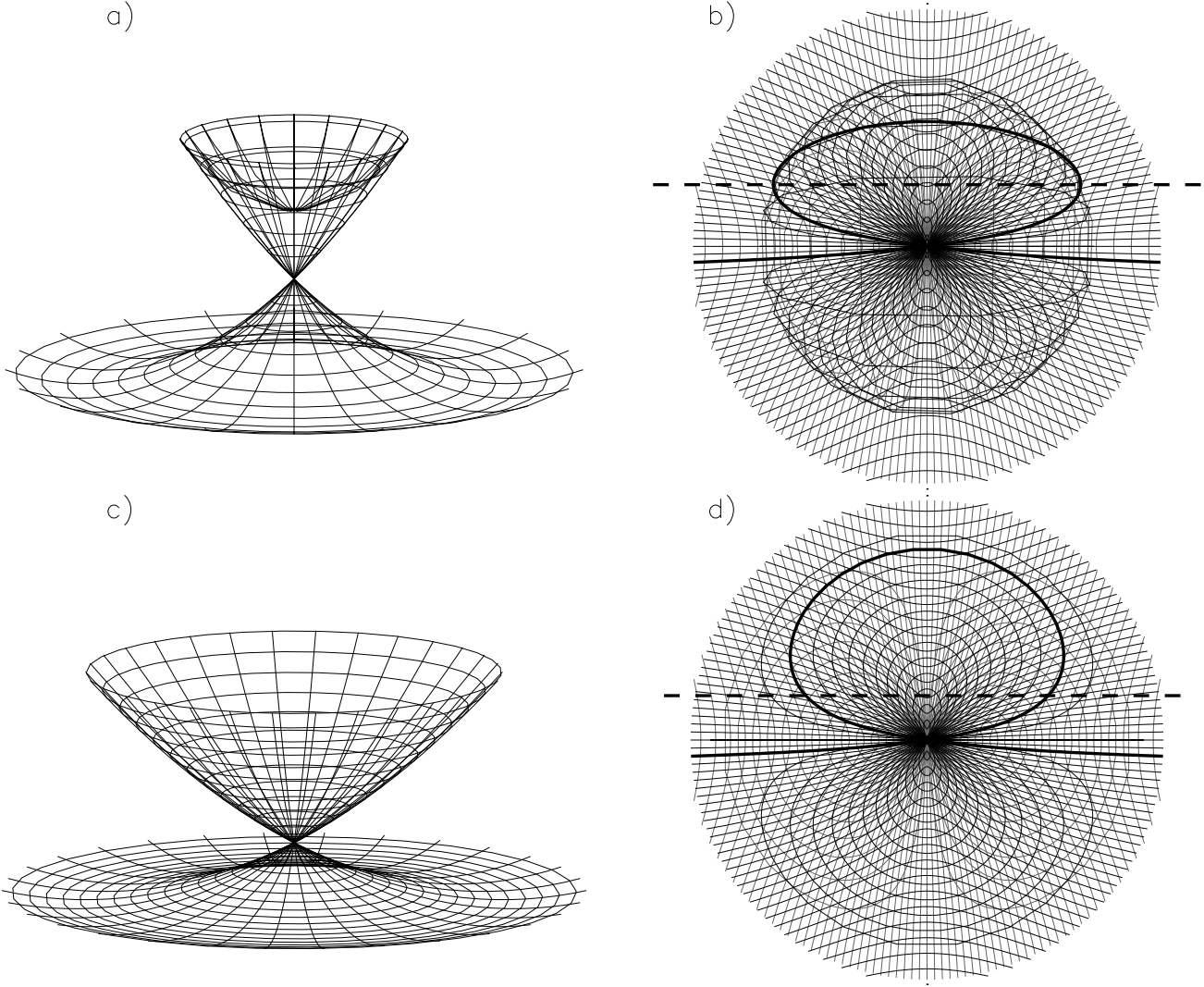,width=20cm}
\vspace{0cm}

\hfill      \parbox[]{16.5cm}{\caption[]{The sky sheet for the
isothermal sphere. Panel a) displays the case of a softened
isothermal sphere with a core radius $\theta_c = 0.2~\theta_E$.
The horizontal axes are the coordinates in the source plane $(\beta_x,
\beta_y)$. The vertical axis is taken proportional to the time delay
associated to each image. The number of images of a source is given by
the times that a vertical line in the source location intersects the
sky sheet. The order of arrival of a light pulse from the source is
proportional to the $z$ coordinate and hence the image with smaller $z$
will show the pulse first. Panel b) shows a projection of case a) into
the $(\beta_x, \beta_y)$ plane. The thick solid line displays the points
observed in the source plane when we move the observation direction
 along a straight line. Panels c) and d) are similar to a) and b) but 
for the case of a singular isothermal sphere. Notice that the upper
cap has disappeared in the singular case and the sky sheet is opened}
\label{skysheet}}%
    \end{figure*}

Figure~\ref{skysheet}a shows this sky sheet for the case of the
softened isothermal sphere. In this case a circularly symmetric fold
develops, which looks like a blob in the surface. The horizontal
coordinates are $(\beta_x, \beta_y$), so that the number of 
intersections of the
vertical lines with the surface give the number of images in the
observer's sky.
Clearly for point
sources located in the region covered by the blob, three images 
will be seen at
three different positions in the sky, while for sources
located outside the blob just one image will be seen. 
Figure~\ref{skysheet}b 
shows the projection of the sky sheet in the $(\beta_x,
\beta_y)$ plane, i.e. as seen from above. The amplification of a
given image is related to how much this surface is stretched at the
point corresponding to the image (here we started with a rectangular
grid so that the stretching  can be visualized, while in 
Figure~\ref{skysheet}a  and \ref{skysheet}c
we used a polar grid to respect the symmetry of the lens). 
If the surface is very stretched, the
magnification is very low, while if it is contracted, the
magnification is high. The circle corresponding to the border of the
blob has then divergent magnification and 
corresponds to a caustic. When a source crosses it, a pair
of new images appears in a different region of the sky, corresponding
to the intersection of the vertical from the source position
$(\beta_x,\beta_y$) with the blob. This is a fold
caustic according to the catastrophe theory classification 
\cite{be76,po78}, and we see that it
actually corresponds to a fold in the sky sheet. The other singular
point is the central one, corresponding to the place where the blob
formed. A source located there (i.e. perfectly aligned with the lens)
will have as images all the points of the circle (i.e. the Einstein
ring) that contracted to
that point. This is a point-like caustic,
which can appear exclusively for circularly symmetric lens
configurations.

It is interesting to note that if we for instance 
point a telescope at a fixed
latitude and move it from left to right along a meridian through a
region of the sky where an isothermal lens is located, we are first
seeing points in the source plane from left to right. When we arrive
to the fold caustic, we start moving in the source plane from right to
left as we move the telescope to the right, and the direction reverses
again when we arrive to the other fold caustic. Thus, we cross the
blob region three times, first to the right, then to the left and
again to the right, as it is clear following the thick line in 
Figure \ref{skysheet}b.  

\begin{figure}
\vspace{0cm}
\hspace{0cm}\psfig{figure=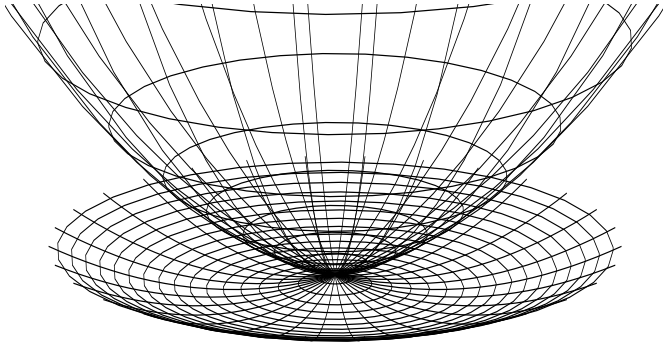,angle=-90,width=8.8cm}
\vspace{0cm}
\hfill      \parbox[]{8cm}{\caption[]{Sky sheet for the 
point-like lens}\label{point}}%
\end{figure}

Figure~\ref{skysheet}c and \ref{skysheet}d show the sky sheet for 
the case of the singular
isothermal sphere. Here, the sky sheet develops a cut in the blob,
transforming  the blob into a cone, and the most demagnified central 
image disappears. Thus,
a source can have just one or two images in this case. The second
image appears or disappears as the source crosses the cut in the sky
sheet and has its parity inverted (for instance, it is easy to see
from the blob picture \ref{skysheet}c which was drawn with polar 
coordinates, that
the $\hat\theta$ unit vector will reverse orientation when projected
into the source plane if it is inside the cone, while the unit vector
in the radial direction will not be reversed, and hence the parity has
to change).  

For the point-like mass the deflection angle formally 
diverges as we approach
the lens direction, thus in this case there is also a cut in the sky
sheet, but the border is stretched to infinity\footnote{Clearly it has
no meaning to consider impact parameters smaller than the radius of
the lens. If the lens is a Schwarzschild black hole, 
a sequence of images is expected to 
appear on both sides of the optical axis, corresponding to light 
rays that go around the lens several times
\cite{vi99}.}, as shown in Figure~\ref{point}. Hence, in this case
there are always two images for any source position (except for
perfect alignment, where the Einstein ring is the 
image). The principal image is  always magnified, 
while the other one can become very  demagnified, what 
is reflected here in that the conical surface becomes very 
stretched as we move out from the lens position.

\begin{figure*}
\vspace{0cm}
\hspace{0cm}\psfig{file=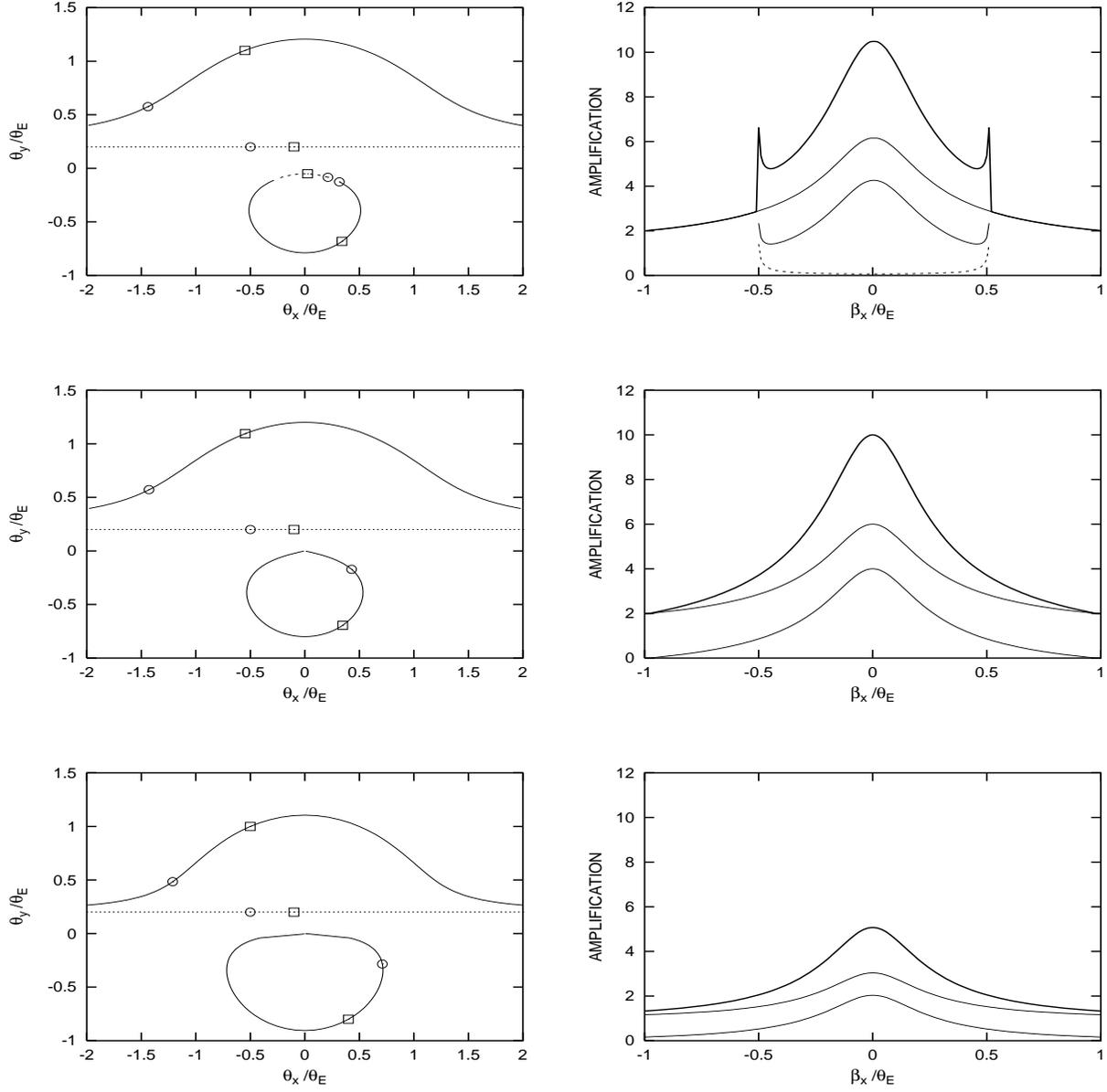,width=16.5cm}
\vspace{0cm}
\hfill      \parbox[]{16.5cm}{\caption[]{Left panels: position of the
images for a source moving along a straight line (dotted) 
with impact parameter
$0.2~\theta_E$ for the softened isothermal sphere (top), the singular
isothermal sphere (middle) and the point-like lens (bottom). The
circles and squares identify the positions of the source and images
for two different times. Right panels: amplification of the individual
images (thin lines) as the source crosses the lens. The
thick lines correspond to the sum of the amplifications.}
\label{ampli}}
\end{figure*}

The magnification of the images is given, in the case of circular
symmetry considered, by $A_i=\left|
\frac{\theta_i}{\beta}{\partial
\theta_i\over\partial\beta}\right|$. 
The resulting expressions for the singular
cases are well known, and for the non--singular case we have found the
amplifications using the solutions $\theta_i(\beta)$ of
eq.~(\ref{poli}). 

In Figure~\ref{ampli} we display illustrative examples for 
each of the three
models. In the left panels we show the image locations (the axes are
$(\theta_x, \theta_y$), measured in units of the corresponding Einstein
radii)  for a source located  at $\beta_y=0.2~\theta_E$ and with varying
$\beta_x$, i.e. moving it along the dotted horizontal line.
Two particular source positions and their images are identified with
an asterisk and a square. From top to bottom the Figures correspond to
the softened isothermal sphere, the singular isothermal sphere and the
point-like mass. This
corresponds to a source crossing the source plane along the dashed
line in Figures~\ref{skysheet}b and \ref{skysheet}d.

   \begin{figure*}
\vspace{0cm}
\hspace{-3cm}\psfig{figure=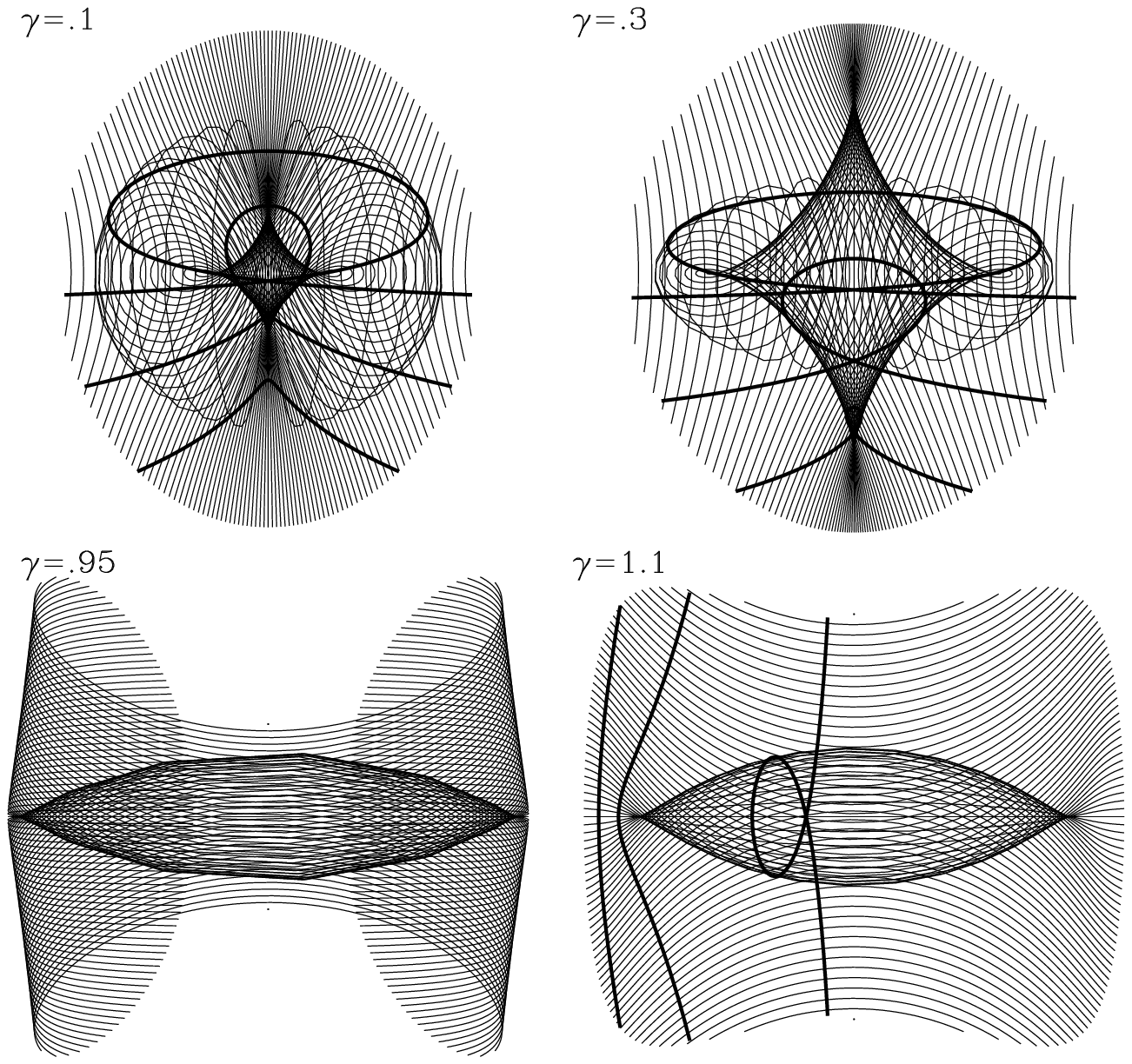,width=20cm}
\vspace{0cm}

\hfill      \parbox[]{16.5cm}{\caption[]{The sky sheet for a singular
isothermal sphere with $\theta_c = 0.2~\theta_E$ in the presence of
an external shear corresponding to different values of $\gamma$.
              }\label{shear}}%
    \end{figure*}

In the right panels of Figure~\ref{ampli} 
we show for the same systems the corresponding
magnifications $A_i$  as a function of $\beta_x$ (measured in units of
$\theta_E$). The thin lines correspond to the individual images, while
the thick lines indicate the sum of the amplifications,
 which is the relevant quantity when the images are unresolved. 
 Notice that in the non singular case, the fainter image
(dashed line in the first panel in Figure~\ref{ampli}) 
is the one closest to the optical axis. Its
magnification would decrease with decreasing core size $\theta_c$, 
becoming  zero  in the singular limit $\theta_c\to 0$ (this explains why 
only two images are seen in the singular cases, the two lower panels in
Figure~\ref{ampli}). 
Due to its significant demagnification, this 
third image is usually missed in the
observations of multiple images of quasars. We also see that besides
the high magnifications of the images for small $\beta$, there are now
also peaks of high magnification associated to the formation or
disappearance of the image pairs at caustic crossing, something that
is not observed in the singular cases.

\subsection{Distorting the point-like singularity}

In the three examples illustrated the lens is circularly symmetric and
hence they all have the point-like caustic in the center, 
whose image is the
Einstein ring. Any departure from this lens symmetry will make the
point-like caustic to distort to several fold caustics merging in
pairs at cusps. For example, we could add to the smooth isothermal
sphere an external shear $\gamma$ and convergence $\kappa$ 
to  model the local effect of the environment around a lensing
galaxy or an elliptical lens \cite{ch84,ko87}).  This effect can be
described by the first terms of the Taylor expansion of the projected 
gravitational potential, which in the principal axes system of the 
shear is
\begin{equation}
\Psi(\theta_x,\theta_y)=\frac{\kappa}{2}(\theta_x^2+\theta_y^2)+
\frac{\gamma}{2}(\theta_x^2-\theta_y^2),
\end{equation}
A non-vanishing shear then clearly 
breaks the circular symmetry of the lens.
For this case the  lens equation can be written as
\begin{eqnarray}
\beta_x & = & \theta_x \left(1 - \kappa - \gamma - \frac{D_{ls}}{D_{os}}
\frac{4 \pi \sigma^2}{c^2}\frac{1}{\sqrt{\theta^2
+\theta_c^2}}\right),\label{betx}\\
\beta_y & = & \theta_y \left(1 - \kappa + \gamma - \frac{D_{ls}}{D_{os}}
\frac{4 \pi \sigma^2}{c^2}\frac{1}{\sqrt{\theta^2
+\theta_c^2}}\right)\label{bety}.
\end{eqnarray}

We show in Figure~\ref{shear}  the sky sheet for a set of values of
$\gamma$ and $\kappa=0$. For small values of $|\gamma|$, such that
$|\kappa \pm \gamma| < 1$, the central point-like caustic is deformed
to four fold caustics ending in four cusps. This can be understood as
follows. In the circularly symmetric case, the Einstein ring
was contracted to a point in the sky sheet, the point of contact of
the blob with the rest of the surface. When we add a shear, there are
for instance 
two different circles that will become shrinked to segments, one in the
$\beta_x$ direction (with radius equal to the value of $\theta$ giving 
$\beta_y=0$ in eq.~(\ref{betx})),  and the other in the $\beta_y$ 
direction (with radius equal to the value of
$\theta$ giving $\beta_x=0$ in eq.~ (\ref{bety})).
When identifying a circle with a segment, the interior of the circle
will give rise to a blob, whose border will be a fold caustic. The
endpoints of the above mentioned segments will correspond to the
merging of two folds into a cusp. 
Depending on its location, a source can
have one, three or five images. We have to count how many times the
point at which the source is located is covered by the sky
sheet. Points that are inside the oblate caustic (in the first two
panels in Figure~\ref{shear}) get a pair of
additional images and those that are inside the diamond shaped caustic
get  another additional pair.

If the value of the shear is increased, the structure of the caustics 
changes. The oblate caustic (pure fold) shrinks inside the diamond
caustic, taking with it the pair of cusps which were in the horizontal
direction (third panel of
Figure~\ref{shear}). In this process the diamond shaped caustic has lost
a pair of cusps and hence also appears as two folds merging into two cusps
(this pair of cusps are far away in the
vertical direction and thus are not displayed in the
figure)\footnote{Let us notice that the folds and cusps that appeared
in the sky sheet figures are the two types of catastrophes of two
dimensional systems.  
One may also identify some higher order catastrophes by considering
in Figure~\ref{shear} the shear as the third parameter. The transition
discussed between the second and third panels corresponds to an
hyperbolic umbilic.}.  

\begin{figure}
\vspace{0cm}
\hspace{0cm}\psfig{figure=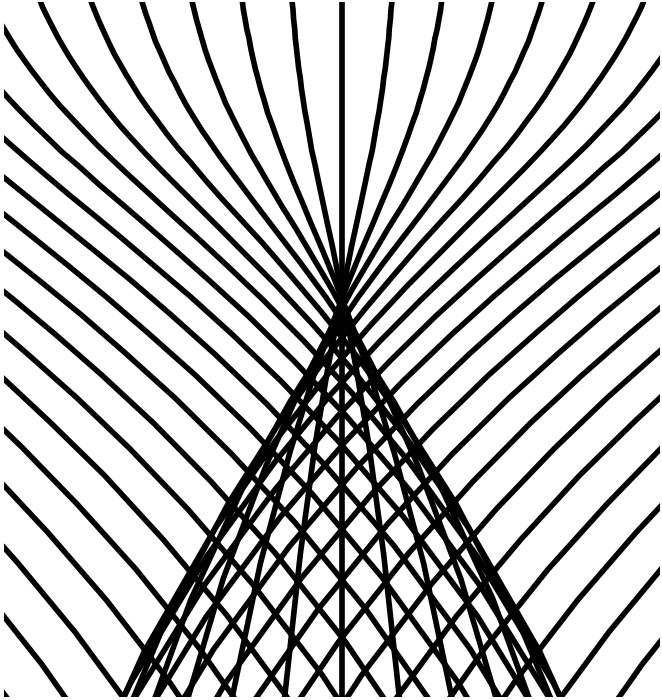,width=8.8cm}
\vspace{0cm}
\hfill      \parbox[]{8cm}{\caption[]{Close view of a cusp: two
caustics merge and the sky sheet unfolds.
              }\label{cusp}}%
\end{figure}
   \begin{figure*}
\vspace{0cm}
\hspace{0cm}\psfig{figure=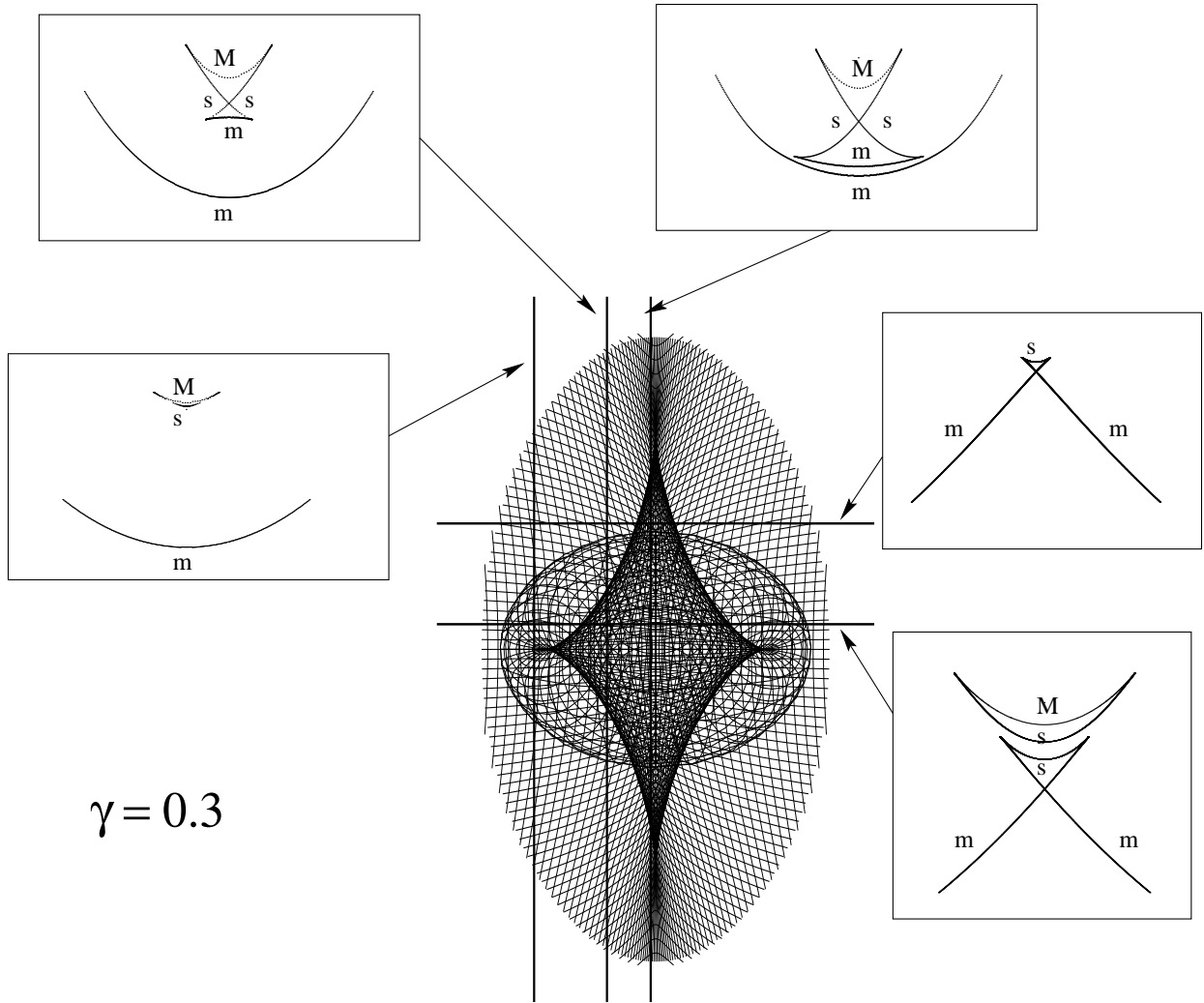,width=16.5cm}
\vspace{0cm}

\hfill      \parbox[]{16.5cm}{\caption[]{Different cuts of the sky sheet
for the isothermal sphere with shear $\gamma = 0.3$. The $z$ coordinate
is proportional to the time delay associated to each image and gives
the order of arrival of a pulse. In each line we indicate if the
images along it correspond to a minimum (m), a saddle point (s) or a
maximum (M) of the time delay function. 
    }\label{cuts}}%
    \end{figure*}

When $
|\gamma| > 1$, the vertical fold disappears (it is stretched to
infinity) and only the horizontal fold shown in the last panel of 
Figure~\ref{shear} remains.
In this case there are just two folds merging into two cusps (lips
singularity).  A source inside the
folds has three images, while one outside has one image. When a source
crosses this lens in a direction transverse to the folds, a pair of
new images adds to the original one in a different place of the sky
when the source crosses the first folding. One of the new images is
inverted, while the other is not. As the source continues
moving, the inverted  image approaches the original image,  merges
with it when the source crosses the second caustic and they disappear,  
leaving as
single image of the source the one which was created in the
crossing of the first fold. When the source enters the folded region
through the cusp, the new pair of images appears in the same 
position as the
original image, and then they separate from each other.
We show in Figure~\ref{cusp}  a close view of a cusp. 
It is clear how the two caustics, corresponding to the
folds in the sky sheet, end in the cusp and the sky sheet surface
unfolds there.

To better appreciate the folds in Figure~\ref{shear} and the time
delay structure of the images,  we present in Figure~\ref{cuts}
different vertical cuts of the second panel in Figure~\ref{shear}. 
The boxes show the sky
sheet profiles along the indicated cuts which allow the visualization
of the folds. We have also indicated the nature of the different
images along each sheet, i.e. whether they are minima ($m$), saddle
points ($s$) or maxima $(M)$ of the time delay function. These
properties change at each fold of the sheet. The pair of images
created in a fold can be a saddle point and a maximum or a saddle point
and a minimum. The saddle points correspond to the images with
inverted parities. The vertical coordinate  we have chosen clearly allows
a simple characterization of the extrema just by inspection of these
cuts and knowing that far from the lens the single images are minima
of the time delay\footnote{except for very large shear, e.g. 
in the last panel
of Figure~\ref{shear}, for which the images outside the blob are saddle
points.}.  If the lens were to become singular, the sky sheet would be
qualitatively similar except that the upper surface (the maximum)
would  be absent, leaving a hole in the sky sheet.

\section{Binary lenses}

Another configuration in which the circular symmetry is broken is when
a second lens is present in the proximity of the original one.
This configuration is of relevance for the study of microlensing by
binary stars or planetary systems.
When the lens is formed by a binary system, the deflection angle of a
given light ray is given by the superposition of the deflections
produced by the two individual lenses. If the lenses, that we take as
softened isothermal spheres with dispersion velocities $\sigma_1$ and
$\sigma_2$, are located in the lens plane forming angles
$\vec\theta_1$ and $\vec\theta_2$ with the chosen optical axis, then
the lens equation can be written as
\begin{equation}
\vec\beta  =  \vec\theta - \mu_1 (\vec\theta-\vec\theta_1)
\sqrt{\frac{\theta_E^2+\theta_c^2}{|\vec\theta-\vec\theta_1|^2
+ \theta_c^2}} - \mu_2 (\vec\theta-\vec\theta_2)
\sqrt{\frac{\theta_E^2+\theta_c^2}{|\vec\theta-\vec\theta_2|^2
+ \theta_c^2}}
\end{equation}
where $\mu_1\equiv \sigma_1^2/\sigma^2$, $\mu_2 \equiv
\sigma_2^2/\sigma^2$, with $\sigma^2=\sigma_1^2+\sigma_2^2$, and
$\theta_E$ refers to the Einstein angle of a single lens with
velocity dispersion $\sigma$ (eq. \ref{einsis})
   \begin{figure*}
\vspace{0cm}
\hspace{-2cm}\psfig{figure=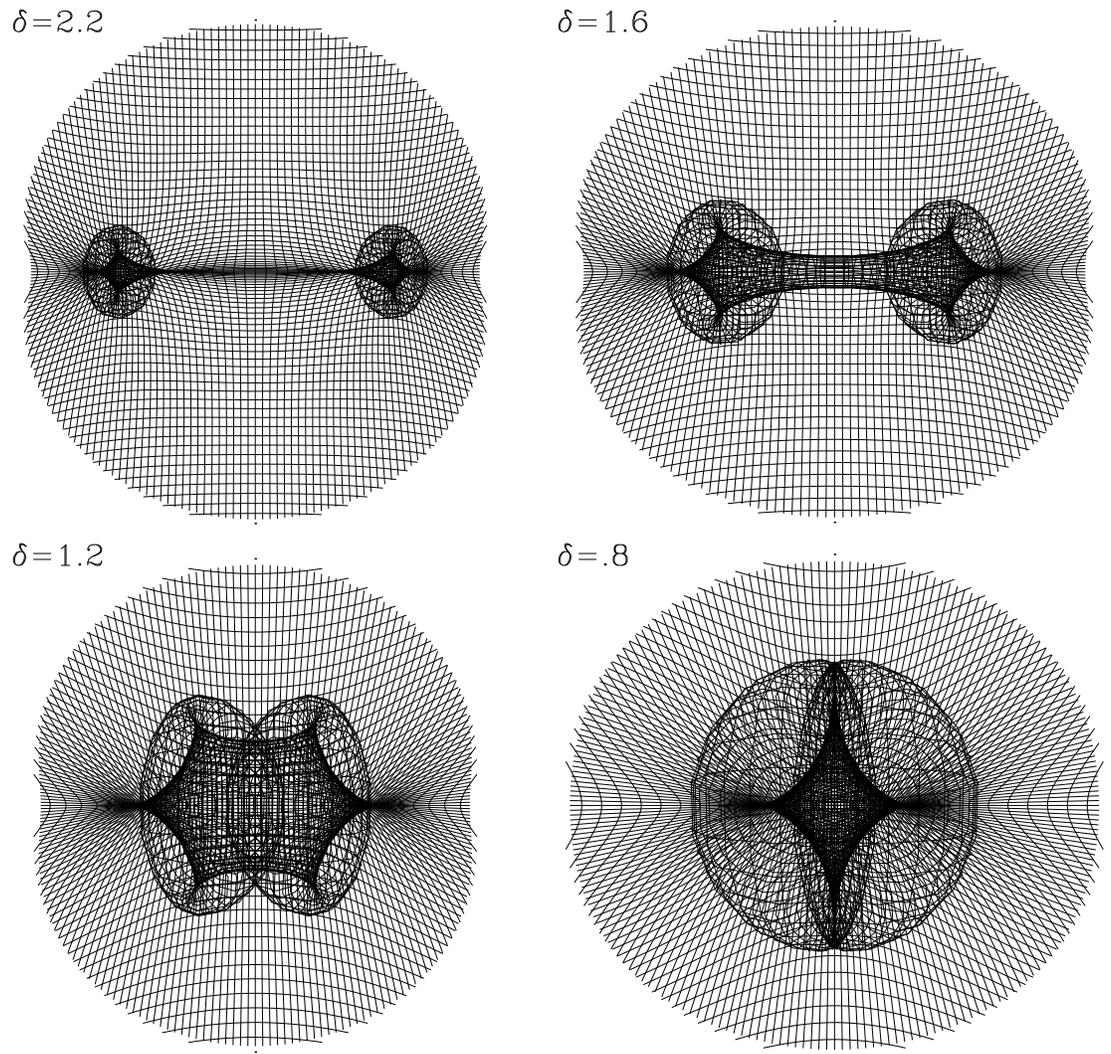,width=20cm}
\vspace{0cm}

\hfill      \parbox[]{16.5cm}{\caption[]{The sky sheet for a binary lens
composed of two identical softened isothermal spheres for different
values of separation between the lenses.
    }\label{bin}}%
    \end{figure*}

The caustics for the case of  
two equal point-like lenses have been studied 
in detail by Schneider and Weiss \cite{sc86}. 
An extension to more general geometries can
be found in Refs. \cite{er93} and 
\cite{do99}. A numerical study of the extension to an ensemble of
compact objects can be found in ref. \cite{ka89}.

We show in Figure~\ref{bin} the sky sheet for the case of equal lenses 
($\sigma_1 = \sigma_2$) and an ensemble of values for the 
lens separation $\delta \equiv |\vec\theta_1-\vec\theta_2|/\theta_E$. 
We took for both lenses a core radius $\theta_c = 0.2~\theta_E$.
As the distribution of matter is smooth, there are no cuts in the
sky sheet and the additional images appear in pairs (there is an odd
number of them for every source position).
If the distance between the lenses is large, 
the caustics associated to each lens are
separated. Each one has a pure fold caustic and a diamond shaped
caustic with four cusps.
A source can have one, three or five images depending on whether it is
located outside the caustics, inside either 
the fold or the diamond caustic,
or inside both of them (first panel). As the lenses
get closer, the diamond shaped caustics become more elongated, until
they merge and form a six cusps caustic as shown in the second panel
\footnote{This merging corresponds to a beak-to-beak transition.}.
For even closer lenses, the two pure fold caustics (the blobs) 
intersect and a region
with seven images develops in the center (third panel). The fourth
panel shows that as the lenses continue to approach each other
the six cusps caustic shrinks to a four cusps caustic (actually, in an
intermediate stage two small triangles are formed at the top and
bottom of the diamond caustic in beak-to-beak singularities, and they
shrink and disappear in a higher order catastrophe, leading to a single
pure fold caustic containing a vertical lips singularity. This last
shrinks and disappears for decreasing $\delta$, and the single lens
caustics are recovered in the limit $\delta\to 0$).

   \begin{figure*}
\vspace{0cm}
\hspace{-2cm}\psfig{figure=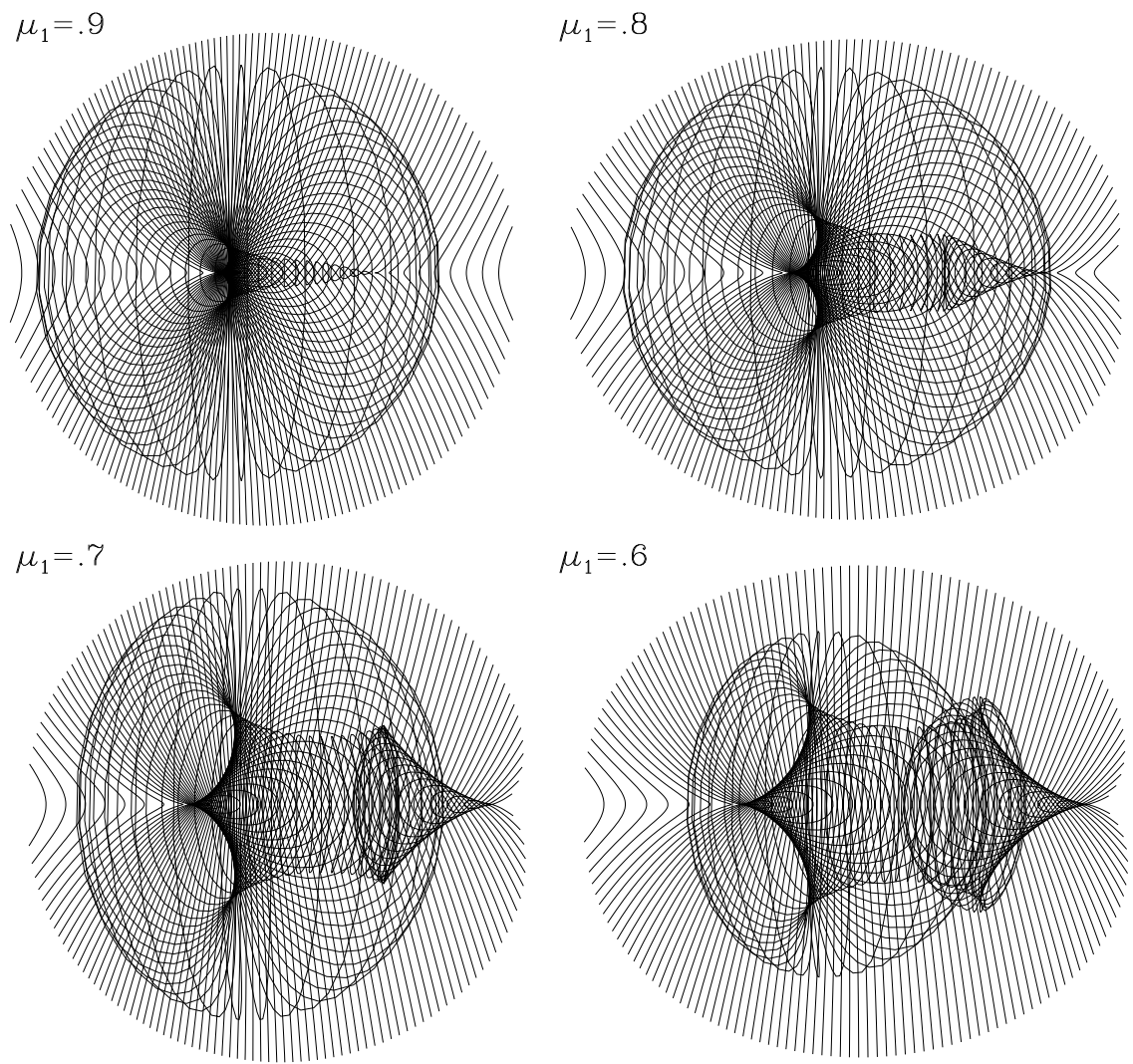,width=20cm}
\vspace{0cm}

\hfill      \parbox[]{16.5cm}{\caption[]{The sky sheet for a binary lens
composed of two softened isothermal spheres separated by a distance
$\delta=1.2$ for different values of the mass ratio $\mu_1$.
    }\label{dm}}%
    \end{figure*}

It is straighforward to generalize this to an asymmetric case.
As an example we show in Figure~\ref{dm} the sky sheet for a
binary lens system with different masses. We take two softened isothermal
spheres with core radius given by $\theta_c= 0.2~\theta_E$.
The distance between the lenses is
fixed at $\delta=1.2$, and the sky sheet is shown for different values
of the parameter $\mu_1$. If one of the masses largely dominates the
system, the sky sheet resembles the single lens case, but instead of 
the central point like caustic there is diamond shaped caustic. As the
second mass becomes more important a new vertical fold  starts
to develop inside the diamond. This fold grows with the mass of the
second lens and eventually cross the border of the diamond,
transforming the diamond in a six cusps caustic and giving rise to
a new pure fold caustic associated to the second lens (this transition
is associated with two hyperbolic umbilics).

\section{Conclusions}

Some basic properties of simple gravitational lensing systems have
been displayed using a graphical approach. This consists in
projecting a regular grid in the image (observers) 
plane to the source plane using
the lens equation. In the strong lensing regime, the mapping is not
one to one, thus the regular grid becomes folded in the
source plane. 

This construction is straightforward and has several
useful properties. In the first place, it gives a clear picture of
which regions of the sky are seen when we look along directions close
to a lensing system. The location of the caustics corresponds to
the folds in the sky sheet. It is also easy to
determine the number of images of a given source by counting the
number of times that the sky sheet covers that position. The
stretching of the sky sheet surface gives an idea of the magnification
of each image.

The sky sheet plots provide also a simple pictorial view of features
like the formation of additional images in pairs when the source
crosses a caustic, of the parity of the images,
or the fact that new images appear close to the
original one when the source crosses a cusp.
 
Finally, the choice of the $z$ coordinate as proportional to the time
delay associated to each image allows to display the ordering of arrival
of any luminosity change of the source along the different image
directions, and also to characterize the kind of extrema of the time
delay function associated with each image.

\bigskip\bigskip

Work partially supported by CONICET, Fundaci\'on Antorchas and Agencia
Nacional de Promoci\'on Cient\'\i fica y Tecnol\'ogica. We thank the
CERN Theory Division for hospitality while part of this work was being
done.

\end{document}